\documentstyle[12pt,bezier]{article}
\begin{document}
\baselineskip 18pt
\def\today{\ifcase\month\or
 January\or February\or March\or April\or May\or June\or
 July\or August\or September\or October\or November\or December\fi
 \space\number\day, \number\year}
\def\thebibliography#1{\section*{References\markboth
 {References}{References}}\list
 {[\arabic{enumi}]}{\settowidth\labelwidth{[#1]}
 \leftmargin\labelwidth
 \advance\leftmargin\labelsep
 \usecounter{enumi}}
 \def\newblock{\hskip .11em plus .33em minus .07em}
 \sloppy
 \sfcode`\.=1000\relax}
\let\endthebibliography=\endlist
\def\epem{e^+e^-}
\def\gsim{{\mathop >\limits_\sim}}
\def\lsim{{\mathop <\limits_\sim}}
\def\r2{\sqrt 2}
\def\tb{\tan\beta}
\def\v#1{v_#1}
\def\wi{\omega_i}
\def\wj{\omega_j}
\def\wk{\omega_k}
\def\xi{\chi_i}
\def\xj{\chi_j}
\def\xk{\chi_k}
\def\m#1{{\tilde m}_#1}
\def\mH{m_H}
\def\mw#1{m_{\omega #1}}
\def\mx#1{m_{\chi #1}}
\def\mwi{m_{\omega i}}
\def\mxj{m_{\chi j}}
\def\ri{\rm i}
\def\rj{\rm j}
\def\rk{\rm k}
\begin{titlepage}
\hspace*{10.0cm}ICRR-Report-399-97-22
   
\hspace*{10.0cm}OCHA-PP-82

\hspace*{10.0cm}kure-pp/97-02
\  \
\vskip 0.5 true cm 
\begin{center}
{\large {\bf $CP$-odd Anomalous $W$-boson Couplings  }}  
{\large {\bf from Supersymmety }}
\vskip 2.0 true cm
\renewcommand{\thefootnote}
{\fnsymbol{footnote}}
Minako Kitahara$^1$\footnote{
Present address:  Sendai Research Laboratory, Kokusai Electric Co. LTD., 
Sendai 981-32, Japan.}, 
Miho Marui$^2$, Noriyuki Oshimo$^3$, 
Tomomi Saito$^1$, and \\
Akio Sugamoto$^1$ 
\\
\vskip 0.5 true cm 
{\it $^1$Department of Physics {\rm and} 
Graduate School of Humanities and Sciences }  \\
{\it Ochanomizu University, Otsuka 2-1-1, 
Bunkyo-ku, Tokyo 112, Japan}  \\
{\it $^2$Faculty of Social Information Science}  \\
{\it Kure University, Gouhara 2411-26, Kure, Hiroshima 724-07, Japan}  \\
{\it $^3$Institute for Cosmic Ray Research}  \\
{\it University of Tokyo, Midori-cho 3-2-1, 
Tanashi, Tokyo 188, Japan}  \\
\end{center}

\vskip 3.0 true cm

\centerline{\bf Abstract}
\medskip
     The supersymmetric standard model contains a new 
$CP$-violating phase in the mass matrices for charginos and 
neutralinos, which could induce $CP$-odd anomalous 
couplings for the $WWZ$ and $WW\gamma$ vertices at the one-loop level.   
We study these couplings, 
paying attention to the model-parameter and $q^2$ dependencies.  
It is shown that the $CP$-odd form factors could have values 
of order $10^{-3}-10^{-4}$, which  
are much larger than those predicted by the standard model. 
We also discuss the possibility of examining these form 
factors in experiments.  

\medskip

\end{titlepage}

\newpage 
\section{Introduction} 

     The study of trilinear gauge-boson vertices $WWZ$ and 
$WW\gamma$ is one of the main subjects for experiments  
at LEPII or near-future $\epem$ colliders.  Their precise 
measurements enable to examine the standard model (SM) in 
which the vertices are determined uniquely.  Possible discrepancies 
between the experimental results and the SM predictions  
would imply the existence of physics beyond 
the SM \cite{hagiwara,schild}.    
Various theoretical analyses therefore have been made 
on the vertices in the SM \cite{peskin} and in its extensions, 
such as the two-Higgs-doublet model \cite{rizzo}, the model 
with Majorana neutrinos \cite{marui}, and the supersymmetric 
model \cite{lahanas}, in particular for $CP$-conserving 
couplings.  

     In this paper we study the trilinear gauge-boson vertices 
focusing on $CP$ violation within the framework of the 
supersymmetric standard model (SSM) \cite{ssm}.  This model contains 
new sources of $CP$ violation as well as the standard 
Kobayashi-Maskawa mechanism.  As a result, the $W$ and $Z$ bosons 
have $CP$-violating interactions with supersymmetric particles \cite{oshimo}.  
These interactions generate $CP$-violating 
couplings for $WWZ$ and $WW\gamma$ at the one-loop level 
\cite{chang, kadoyoshi}.  
Since the SM does not predict $CP$ violation for the vertices 
at the tree level nor the one-loop level, observation of 
$CP$-violating phenomena arising from these vertices could immediately indicate 
the existence of physics other than the SM \cite{wwcp}.  
The SSM gives radiative corrections also to $CP$-even couplings 
for the vertices \cite{bilchak}, 
although they are at most of the same order of magnitude as the 
SM predictions  \cite{lahanas}.   

     The new sources of $CP$ violation of the SSM can also give 
contributions to the electric dipole moments (EDMs) of the 
neutron and the electron through one-loop diagrams 
mediated by the charginos, neutralinos, or gluinos, together 
with the squarks or sleptons.  
For wide ranges of SSM parameters the magnitudes of the induced EDMs 
could be around or even larger than their present experimental upper bounds, 
thus providing non-trivial constraints on the SSM.  We assume that 
$CP$-violating phases intrinsic in the SSM have a natural 
magnitude of order unity, 
since there is no convincing reason which suppresses them.   
Then, the masses of the squarks and sleptons 
are constrained from the EDMs to be larger than 1 TeV, 
while the charginos and neutralinos 
could have masses of order of 100 GeV \cite{edm}.  

     The one-loop diagrams which induce the $CP$-violating 
couplings for $WWZ$ and $WW\gamma$ in the SSM 
could be mediated by various supersymmetric particles.  
However, if the new $CP$-violating phases are not suppressed, 
the squarks and sleptons have to be 
heavy and thus the diagrams with these particles 
can be neglected.  
Sizable contributions to the couplings could only be generated through 
the diagrams mediated by the charginos and neutralinos shown 
in Fig. 1, on which our analyses are concentrated throughout 
this paper.      

    This paper is organized as follows.  In sect. 2 the $CP$-violating 
interactions in the SSM are briefly summarized.  In sect. 3 we obtain 
the $CP$-odd form factors for the $WWZ$ and $WW\gamma$ vertices  
and make numerical analyses in detail.   
The possibility of detecting the $CP$-odd couplings 
is discussed in sect. 4.  

\section{Model}

     The $CP$-odd couplings for the $W$ bosons are induced by the 
interactions of charginos $\wi$ and neutralinos $\xj$, 
the charged and neutral mass 
eigenstates of higgsinos and SU(2)$\times$U(1) gauginos.   
Their mass matrices are given by 
\begin{eqnarray}
    M^- &=& \left(\matrix{\m2 & -g\v1^*/\r2 \cr
                       -g\v2^*/\r2 & \mH}        \right), 
\label{eq1} \\
   M^0 &=& \left(\matrix{\m1 &  0  & g'\v1^*/2 & -g'\v2^*/2 \cr
                         0  & \m2 & -g\v1^*/2 &   g\v2^*/2 \cr
                       g'\v1^*/2 & -g\v1^*/2 &   0  & -\mH \cr
                      -g'\v2^*/2 &  g\v2^*/2 & -\mH &   0}
           \right), 
\label{eq2}
\end{eqnarray}
where 
$\v1$ and $\v2$ are the vacuum expectation values
of the Higgs bosons; $m_H$ is the mass parameter in the bilinear term 
of the Higgs superfields in superpotential;  and $\tilde{m_2}$ 
and $\m1$ are the masses of SU(2) and U(1) gauginos, respectively, 
appearing in the soft supersymmetry-breaking terms.  
In general, these parameters have complex values.  
Although there is some freedom of redefining the phases 
of the fields, all the complex phases cannot be rotated away.   
The mass matrices $M^-$ and $M^0$ are diagonalized 
to give mass eigenstates: 
\begin{eqnarray}
      C_R^\dagger M^-C_L &=& {\rm diag}(\mw1, \mw2) \quad 
                       (\mw1 <\mw2 ),    
\label{eq3} \\
N^tM^0N &=& {\rm diag}(\mx1, \mx2, \mx3, \mx4) \quad
                       (\mx1<\mx2<\mx3<\mx4), 
\label{eq4}
\end{eqnarray}
where $C_R$, $C_L$, and $N$ are unitary matrices.  

     The complex mass matrices for the charginos and the 
neutralinos lead to $CP$-violating interactions.  
The interaction Lagrangian for the charginos, neutralinos, and $W$ bosons  
are given by  
\begin{equation}
 {\cal L} = \frac{1}{\sqrt{2}}g\overline{\xj}\gamma^\mu               
     \left( G_{Lji}\frac{1-\gamma_5}{2}
                      +G_{Rji}\frac{1+\gamma_5}{2} \right)  
                         \wi W_\mu^\dagger + {\rm h.c.},  
\label{eq5}
\end{equation}
\begin{eqnarray}
 G_{Lji} &=&  \sqrt{2} N_{2j}^\ast C_{L1i}+N_{3j}^\ast C_{L2i},  
\nonumber \\
 G_{Rji} &=&  \sqrt{2} N_{2j} C_{R1i}-N_{4j} C_{R2i}.  
\nonumber 
\end{eqnarray}
The interaction Lagrangian for the charginos and 
the neutralinos with the $Z$ boson is given by  
\begin{eqnarray}
 {\cal L}&=&\frac{1}{\cos\theta_{\rm W}}g
\biggl\{
          \overline{\wi}\gamma^\mu
            \left( F_{Lij}\frac{1-\gamma_5}{2}
               +F_{Rij}\frac{1+\gamma_5}{2} \right) \wj  \nonumber \\
          & & -\frac{1}{4}\overline{\xi}\gamma^\mu
             \left( F_{ij}\frac{1-\gamma_5}{2}
               -F^\ast_{ij}\frac{1+\gamma_5}{2} \right) \xj
\biggr\}Z_\mu ,   
\label{eq6} 
\end{eqnarray}
\begin{eqnarray}
     F_L &=& \left( \matrix{
                     \cos^2\theta_{\rm W}-\frac{1}{2}|C_{L21}|^2 &
                    -\frac{1}{2}C_{L21}^\ast C_{L22} \cr
                    -\frac{1}{2}C_{L22}^\ast C_{L21} &
                     \cos^2\theta_{\rm W}-\frac{1}{2}|C_{L22}|^2 
                           }\right),  
\nonumber \\
     F_R &=& \left( \matrix{
                       \cos^2\theta_{\rm W}-\frac{1}{2}|C_{R21}|^2 &
                      -\frac{1}{2}C_{R21}^\ast C_{R22} \cr
                      -\frac{1}{2}C_{R22}^\ast C_{R21} &
                       \cos^2\theta_{\rm W}-\frac{1}{2}|C_{R22}|^2
                            }\right),  
\nonumber \\
     F_{ij} &=& N_{3i}^\ast N_{3j} - N_{4i}^\ast N_{4j}.  
\nonumber
\end{eqnarray}
The SSM parameters which determine the
interactions in Eqs. (\ref{eq5}) and (\ref{eq6}) are $\v1$,
$\v2$, $\m2$, $\m1$, and $\mH$
appearing in Eqs. (\ref{eq1}) and (\ref{eq2}).  We assume the relation
$\m1=(5/3)\tan^2\theta_W\m2$ suggested by grand unified theories.  
The redefinitions of the fields make it possible without loss of
generality to take all these parameters
except $\mH$ real and positive.  
Then, the $CP$-violating phase is represented by the phase of $\mH$,
which we express as
\begin{equation}
\mH=|\mH|\exp(i\theta).
\label{eq7}
\end{equation}
Since $\v1$ and $\v2$ are related to the $W$-boson mass $M_W$,
independent parameters become $\tb$, $\m2$, $|\mH|$,
and $\theta$, where the ratio of the vacuum expectation values $\v2/\v1$ 
is denoted by $\tb$.

\section{Form factors}

     For the pair production of the $W$ bosons in $\epem$ annihilation 
the trilinear gauge-boson vertex $WWV$, 
$V$ being $Z$ or $\gamma$, can generally be expressed 
in momentum space as \cite{hagiwara}  
\begin{eqnarray}
\Gamma_V^{\nu\lambda\mu}(p, \bar{p}, q) &=& 
       f_{1}^V (p-\bar{p})^{\mu} g^{\nu\lambda} 
     - f_{2}^V (p-\bar{p})^{\mu} q^{\nu} q^{\lambda}/M_{W}^{2} 
     + f_{3}^V ( q^{\nu} g^{\mu\lambda} - q^{\lambda} g^{\mu\nu} ) 
                      \nonumber \\ 
 & & + i f_{4}^V (q^{\nu} g^{\mu\lambda} + q^{\lambda} g^{\mu\nu} ) 
     + i f_{5}^V \varepsilon^{\mu\nu\lambda\rho} (p - \bar{p} )_{\rho} 
     - f_{6}^V \varepsilon^{\mu\nu\lambda\rho} q_{\rho}  
                      \nonumber \\
 & & - f_{7}^V ( p - \bar{p} )^{\mu}\varepsilon^{\nu\lambda\rho\sigma}
                 q_{\rho} (p-\bar{p} )_{\sigma}/M_{W}^{2}, 
\end{eqnarray}
where $q$, $p$ and $\bar p$ are the momenta of the vector bosons
$V_\mu$, $W_\mu^-$, and $W_\mu^+$, respectively. 
Among the seven form factors, $f_1^V$, $f_2^V$, $f_3^V$, and $f_5^V$ 
are $CP$-even and $f_4^V$, $f_6^V$, and $f_7^V$ are $CP$-odd.   
In the SM the form factors $f_1^V$ and $f_3^V$ alone do not vanish 
at the tree level, which holds also in the SSM.  

     The $CP$-odd form factors receive contributions from 
the one-loop diagrams in Fig. 1.  The $WWZ$ couplings are 
generated by Fig. 1(a) and Fig. 1(b), while the 
$WW\gamma$ couplings by Fig. 1(a).  We obtain the form 
factors as follows.  

\noindent
i) The $WWZ$ vertex:
\begin{eqnarray} 
f_4^Z &=& f_{4(a)}^Z+f_{4(b)}^Z,  
\label{eq9} \\
f^Z_{4(a)}
     &=&\frac{-1}{(4\pi)^2}\frac{g^2}{\cos^2\theta_{\rm W}}\frac{1}{2}
         \sum_{i=1}^2 \sum_{j=1}^2 \sum_{k=1}^4    \nonumber \\
     & & \times \biggl\{
              {\rm Im} \left[G_{Lki}G^\ast_{Lkj}F_{Lij}
                       +G_{Rki}G^\ast_{Rkj}F_{Rij} \right]
      \left( -M^2_WA_3 + (m_{\omega j}^2-m_{\omega i}^2)S_3 \right)
                                                   \nonumber \\
     & & {}+  {\rm Im} \left[G_{Lki}G^\ast_{Lkj}F_{Rij}
                       +G_{Rki}G^\ast_{Rkj}F_{Lij} \right]
        m_{\omega i}m_{\omega j}A_1 
                                                   \nonumber \\
     & & {}+  {\rm Im} \left[G_{Lki}G^\ast_{Rkj}F_{Lij}
                       +G_{Rki}G^\ast_{Lkj}F_{Rij} \right]
        m_{\omega j}m_{\chi k} S_2 
                                                   \nonumber \\
     & & {}+  {\rm Im} \left[G_{Lki}G^\ast_{Rkj}F_{Rji}
                       +G_{Rki}G^\ast_{Lkj}F_{Lij} \right]
       \left( -m_{\omega i}m_{\chi k} S_2 \right)
                \biggr\}, \nonumber \\
  f_{4(b)}^Z
     &=&\frac{-1}{(4\pi)^2}\frac{g^2}{\cos^2\theta_{\rm W}}\frac{1}{4} 
         \sum_{i=1}^4 \sum_{j=1}^4 \sum_{k=1}^2    \nonumber \\
     & & \times \biggl\{
 {\rm Im} \left[G_{Lik}G^\ast_{Ljk}F_{ji}
                       -G_{Rik}G^\ast_{Rjk}F^\ast_{ji} \right]
      \left( -M^2_WA_3 + (m_{\chi j}^2-m_{\chi i}^2)S_3 \right)
                                                   \nonumber \\
     & & {}+  {\rm Im} \left[G_{Lik}G^\ast_{Ljk}F^\ast_{ji}
                       -G_{Rik}G^\ast_{Rjk}F_{ji} \right]
      \left( -m_{\chi_i}m_{\chi_j}A_1 \right)
                                                   \nonumber \\
     & & {}+  {\rm Im} \left[G_{Lik}G^\ast_{Rjk}F_{ji}
                       -G_{Rik}G^\ast_{Ljk}F^\ast_{ji} \right]
        m_{\chi j}m_{\omega k}S_2 
                                                   \nonumber \\
     & & {}+  {\rm Im} \left[G_{Lik}G^\ast_{Rjk}F^\ast_{ji}
                       -G_{Rik}G^\ast_{Ljk}F_{ji} \right]
        m_{\chi i}m_{\omega k}S_2 
                \biggr\},       \nonumber \\
f_6^Z &=& f_{6(a)}^Z+f_{6(b)}^Z, 
\label{eq10} \\
f^Z_{6(a)}
     &=&\frac{1}{(4\pi)^2}\frac{g^2}{\cos^2\theta_{\rm W}}\frac{1}{2}
         \sum_{i=1}^2 \sum_{j=1}^2 \sum_{k=1}^4    \nonumber \\
     & & \times \biggl\{
              {\rm Im} \left[G_{Lki}G^\ast_{Lkj}F_{Lij}
                       -G_{Rki}G^\ast_{Rkj}F_{Rij} \right]
\nonumber \\
     & & ~~~~~ \times
       \left( -M^2_W(A_3-2A_2)-2q^2A_4
                  +3(m_{\omega j}^2-m_{\omega i}^2)S_3 \right)
                                                   \nonumber \\
     & & {}+  {\rm Im} \left[G_{Lki}G^\ast_{Lkj}F_{Rij}
                       -G_{Rki}G^\ast_{Rkj}F_{Lij} \right]
        m_{\omega i}m_{\omega j}A_1 
                                                   \nonumber \\
     & & {}+  {\rm Im} \left[G_{Lki}G^\ast_{Rkj}F_{Lij}
                       -G_{Rki}G^\ast_{Lkj}F_{Rij} \right]
        m_{\omega j}m_{\chi k}(A_1-S_1) 
                                                   \nonumber \\
     & & {}+  {\rm Im} \left[G_{Lki}G^\ast_{Rkj}F_{Rij}
                       -G_{Rki}G^\ast_{Lkj}F_{Lij} \right]
        m_{\omega i}m_{\chi k}(-A_1-S_1) 
                \biggr\}, 
\nonumber \\
f_{6(b)}^Z
     &=&\frac{1}{(4\pi)^2}\frac{g^2}{\cos^2\theta_{\rm W}}\frac{1}{4}
         \sum_{i=1}^4 \sum_{j=1}^4 \sum_{k=1}^2    \nonumber \\
     & & \times \biggl\{
              {\rm Im} \left[G_{Lik}G^\ast_{Ljk}F_{ji}
                       +G_{Rik}G^\ast_{Rjk}F^\ast_{ji} \right]
\nonumber \\
     & & ~~~~~ \times 
        \left( M^2_W(A_3-2A_2)+2q^2A_4
           +3(m_{\chi i}^2-m_{\chi j}^2)S_3 \right)
                                                   \nonumber \\
     & & {}+  {\rm Im} \left[G_{Lik}G^\ast_{Ljk}F^\ast_{ji}
                       +G_{Rik}G^\ast_{Rjk}F_{ji} \right]
         m_{\chi i}m_{\chi j}A_1
                                                   \nonumber \\
     & & {}+  {\rm Im} \left[G_{Lik}G^\ast_{Rjk}F_{ji}
                       +G_{Rik}G^\ast_{Ljk}F^\ast_{ji} \right]
        m_{\chi j}m_{\omega k}(-A_1+S_1) 
                                                   \nonumber \\
     & & {}+  {\rm Im} \left[G_{Lik}G^\ast_{Rjk}F^\ast_{ji}
                       +G_{Rik}G^\ast_{Ljk}F_{ji} \right]
        m_{\chi i}m_{\omega k}(-A_1-S_1) 
                \biggr\}, 
\nonumber \\
f^Z_7&=&0,   
\label{eq11} 
\end{eqnarray}
where $S_i$ ($i=1-3$) and $A_i$ ($i=1-4$) stand for 
the functions defined by  
\begin{equation}
\hspace{-2cm}
\left[ S_1,S_2,S_3 \right] \equiv
      \int_0^1\!\!dx\int_0^1\!\!dy\int_0^1\!\!dz
      ~~\frac{\delta (1-x-y-z)}{D(m_i,m_j,m_k)}
      \left[ 1,z,xy \right],\nonumber
\end{equation}
\begin{eqnarray*}
 \lefteqn{ \left[ A_1,A_2,A_3,A_4 \right] \equiv }\\
& & \int_0^1\!\!dx\int_0^1\!\!dy\int_0^1\!\!dz
 ~~\frac{\delta (1-x-y-z)}{D(m_i,m_j,m_k)}
   \left[ x-y,z(x-y),z^2(x-y),xy(x-y) \right],
\end{eqnarray*} 
\begin{eqnarray}
D(m_i,m_j,m_k)\equiv
{}-M^2_W~z(1-z)-q^2xy+m^2_ix+m^2_jy+m^2_kz-i\varepsilon.\nonumber
\end{eqnarray}
For $f^Z_{4(a)}$ and $f^Z_{6(a)}$ the arguments of these functions 
are given by $m_i=\mwi$, $m_j=m_{\omega j}$, and 
$m_k=m_{\chi k}$, 
and for $f^Z_{4(b)}$ and $f^Z_{6(b)}$ given by $m_i=m_{\chi i}$, 
$m_j=\mxj$, and $m_k=m_{\omega k}$.  
The form factors $f_{i(a)}^Z$ and $f_{i(b)}^Z$ $(i=4, 6)$ arise from 
the diagrams in Fig. 1(a) and Fig. 1(b), respectively.  

\noindent
ii) The $WW\gamma$ vertex:  
\begin{eqnarray}
f_{6}^\gamma
     &=&\frac{-1}{(4\pi)^2}g^2 \sum_{i=1}^2 \sum_{k=1}^4 
         {\rm Im} \Bigl[G_{Lki}G^\ast_{Rki}-G^\ast_{Lki}G_{Rki} \Bigr]
         m_{\omega i}m_{\chi k}S_1,
\label{eq12} \\
f_4^\gamma &=& 0,   
\label{eq13} \\
f_7^\gamma &=& 0.   
\label{eq14} 
\end{eqnarray}
The arguments of $S_1$ are given by $m_i=m_j=\mwi$ and $m_k=m_{\chi k}$.  
The nonvanishing contribution to $f^\gamma_6$ only comes from 
Fig. 1(a) with the charginos being the same $i=j$.  

     We now make numerical analyses of the form factors.  The 
numerical computations of one-loop integrals have been carried out 
following the method of \cite{kato}.  
For the SM parameters we fix $\sin^2\theta_W=0.232$, 
$M_Z=M_W/\cos\theta_W=91.2$ GeV, 
and $\alpha_{EM}=1/128.9$.      
In Fig. 2 we show the absolute values of the real and 
imaginary parts of $f_4^Z$ and $f_6^Z$ as functions of 
the absolute value of the higgsino mass parameter $\mH$ 
for four sets of values of $\m2$ and $\tb$ listed in Table 1.  
For the $CP$-violating phase we take $\theta=\pi/4$.  
The value of the momentum-squared for the $Z$ boson 
is set for $\sqrt{q^2}=200$ GeV.  In the 
ranges of smaller values for $|\mH|$ where curves are not drawn, 
the lighter chargino has a mass smaller than 45 GeV, 
which has been ruled out by LEP experiments \cite{PDG}.  
If the masses of the particles which couple to the $Z$ boson 
in a loop diagram  
are near the threshold, $\sqrt{q^2}\simeq m_i+m_j$, 
the contributions of this diagram to the 
form factors are enhanced.  
Below the threshold, the diagram does not generate the 
imaginary parts of the form factors.  
In the discussed parameter ranges, 
there exist several regions which are near the thresholds for 
pairs of charginos or neutralinos.  
This is the reason why the 
form factors complicatedly depend on the parameters.  
The values of $|{\rm Re}(f^Z_6)|$ and $|{\rm Im}(f^Z_6)|$ can become   
around $5\times 10^{-4}$ for $\tb=2$, $\m2=100$ GeV, 
and $|\mH|\sim 100$ GeV.  The magnitudes of 
${\rm Re}(f^Z_4)$ and ${\rm Im}(f^Z_4)$ are smaller than 
those for $f^Z_6$ and at most around $1\times 10^{-4}$.   
Larger values for $\tb$ suppress the form factors. 

     In Fig. 3 the absolute values of the real and imaginary 
parts of $f_6^\gamma$  
are shown for the same parameter values as in Fig. 2.  
The values of $|{\rm Re}(f_6^\gamma)|$ and $|{\rm Im}(f_6^\gamma)|$ 
can become around $5\times 10^{-4}$ for $\tb=2$, $\m2=100$ GeV, 
and $|\mH|\sim 100$ GeV.  As $\tb$ increases, 
$|{\rm Re}(f_6^\gamma)|$ and $|{\rm Im}(f_6^\gamma)|$ decrease.  
Since the photon only couples to the pair of the same charginos, 
the parameter dependence of $f^\gamma_6$ is much simpler 
than $f^Z_4$ or $f^Z_6$.   

     In Figs. 4 and 5 we show the $\sqrt{q^2}$ dependencies of the 
form factors.  Four curves (a), (b), (c), and (d) in Fig. 4 represent 
the absolute values of  
Re($f_4^Z$), Im($f_4^Z$), Re($f_6^Z$), and Im($f_6^Z$), respectively,  and 
two curves (a) and (b) in Fig. 5 the absolute values of 
Re($f_6^\gamma$) and Im($f_6^\gamma$), respectively.  
The parameters are set for $\m2=200$ GeV, $|\mH|=200$ GeV, 
$\tb=2$, and $\theta=\pi/4$.  
These parameter values lead to the masses of the charginos and 
the neutralinos shown in Table 2.   
By the same reason as for Figs. 2 and 3, the form factors $f^Z_4$ 
and $f^Z_6$ depend on $\sqrt{q^2}$ complicatedly, while $f^\gamma_6$ 
does simply.  The magnitudes of ${\rm Re}(f^Z_6)$, ${\rm Im}(f^Z_6)$,  
${\rm Re}(f^\gamma_6)$, and ${\rm Im}(f^\gamma_6)$ 
can become around $2\times 10^{-4}$, though those of 
${\rm Re}(f^Z_4)$ and  ${\rm Im}(f^Z_4)$ are at most $2\times 10^{-5}$.      
If $\m2$ and $|\mH|$ are of order 100 GeV and 
$\tb$ is not much larger than unity, 
in general, there is a region of $\sqrt{q^2}$ where $|f^Z_6|$ or 
$|f^\gamma_6|$ becomes larger than $1\times 10^{-4}$.    

\section{Discussions}

     We have shown that the SSM yields $CP$-odd anomalous couplings 
for the $WWZ$ and $WW\gamma$ vertices at the one-loop level.  
The $CP$-odd form factors could have magnitudes of 
$10^{-3}-10^{-4}$, which are far larger than the SM predictions.  
If some $CP$-violating phenomena originating from the $CP$-odd 
couplings are observed, the SSM would become more promising 
as physics beyond the SM.     

     We now discuss the effects of the $CP$-odd 
form factors on observable quantities.  
The resultant $CP$-violating phenomena occur primarily in 
the pair production of polarized 
$W$ bosons in $\epem$ annihilation,  
$\epem\rightarrow W^+(\bar\lambda)W^-(\lambda)$, 
$\bar\lambda$ and $\lambda$ denoting respectively the helicities 
of $W^+$ and $W^-$.  
Among various combinations for the helicities $(\bar\lambda,\lambda)$, 
the pairs $(+,0)$, $(-,0)$, and $(+,+)$ are $CP$-conjugate 
to the pairs $(0,-)$, $(0,+)$, and $(-,-)$, respectively.  
If $CP$ invariance is conserved, these $CP$-conjugate processes 
have the same cross sections.    
Thus, nonvanishing values for the 
differences of the cross sections 
$\sigma_{+0}-\sigma_{0-}$, 
$\sigma_{-0}-\sigma_{0+}$, and 
$\sigma_{++}-\sigma_{--}$ 
exhibit $CP$ violation.  These differences 
are indeed generated by the imaginary parts of the 
$CP$-odd form factors.  For instance, 
$CP$ violation can be evaluated by an asymmetry 
\begin{equation}
  A_{CP}=\frac{\sigma_{+0}-\sigma_{0-}}{\sigma_{+0}+\sigma_{0-}}, 
\end{equation}
which is given by 
\begin{eqnarray}
A_{CP} &=& 
  \left(-1+\frac{s}{s-M_Z^2}\right)^{-1}
     \left\{2I^\gamma +\frac{s}{s-M_Z^2}(I^\gamma +I^Z) 
     +\left(\frac{s}{s-M_Z^2}\right)^22I^Z\right\},   \nonumber \\
  I^V &=& {\rm Im}(f_4^V) - \frac{{\rm Im}(f_6^V)}{\beta},  
\end{eqnarray}
where $\beta=\sqrt{1-4M_W^2/q^2}$.   
We can see that 
the magnitude of $A_{CP}$ becomes of order of the $CP$-odd form 
factors.  In the SSM such $CP$ asymmetries could thus be of 
order of $10^{-3}-10^{-4}$ in a maximal case.   
On the other hand, the real parts of the $CP$-odd form factors induce  
$T$ violation in the angular distribution of the polarization 
vector {\boldmath $\epsilon$}  
for the $W^+$ or $W^-$ boson, leading to a $T$-odd 
asymmetry \cite{oshimo,kizukuri} 
\begin{equation}
 A_T=\frac{\sigma({\bf (p_-\times p)}\cdot \mbox{\boldmath $\epsilon$} >0)-
    \sigma({\bf (p_-\times p)}\cdot \mbox{\boldmath $\epsilon$} <0)}
         {\sigma({\bf (p_-\times p)}\cdot \mbox{\boldmath $\epsilon$} >0)+
          \sigma({\bf (p_-\times p)}\cdot \mbox{\boldmath $\epsilon$} <0)},  
\end{equation}
where $\bf p_-$ and $\bf p$ denote the momenta of the 
electron and the $W$ boson, respectively.  
The value of $A_T$ becomes also the same order of magnitude as the $CP$-odd 
form factors.  

     The helicity of the $W$ boson affects the 
energy distribution of the particle produced by 
the $W$-boson decay.  
Consequently the $CP$ asymmetry $A_{CP}$ for the $W$-boson pair production 
could be observed as an asymmetry 
between the energy distributions of the particles 
produced from $W^+$ and $W^-$.  Unless the contributions 
of different $CP$-conjugate pairs are canceled, this  
resultant asymmetry would be of the same order of magnitude as $A_{CP}$.  
The $T$-odd asymmetry $A_T$ leads to some $T$-odd asymmetry among  
the particle momenta in the final state \cite{wwcp} 
with the same order of magnitude.  
Assuming maximal $CP$ violation, a total of $10^6-10^8$ 
pairs of $W$ bosons would make it possible to examine 
these asymmetries.    
However, in near-future $\epem$ experiments 
it seems to be difficult to achieve such a number of events \cite{miyamoto}.  

     The form factor $f^\gamma_6$ for the $WW\gamma$ 
vertex could be measured indirectly by the 
EDMs of the neutron and the electron.   
If this form factor has a nonvanishing value, 
these EDMs receive contributions from one-loop diagrams 
generated by SM interactions \cite{marciano}.  
For $f^\gamma_6\sim 10^{-4}$, the neutron EDM is predicted to be 
of order $10^{-26}e$cm \cite{kadoyoshi}, which is smaller than the present 
experimental upper bound by only one order of magnitude.  
The improvement for precision of the experiments is 
expected \cite{yoshiki}, so that the EDMs may 
be able to disclose the $CP$-odd form factor.  
Another possibility of indirect measurements 
is in neutron-nucleus collisions at very 
low energy, where enormous resonance enhancement for 
$P$-violating effects has been observed \cite{masuda}.  
The same enhancement for $T$-violating effects is expected, 
by which $CP$-odd couplings of the neutron and the $Z$ boson 
could be probed very precisely \cite{yamaguchi}.  
Such couplings can be generated at the one-loop level by 
SM interactions, if $CP$-odd couplings for $WWZ$ are nonvanishing.   
If the $CP$-odd couplings of the neutron and the $Z$ boson 
can be explored to the same order of magnitude as the neutron EDM,  
the form factor $f^Z_6$ for $WWZ$ may become detectable.  

\section*{Acknowledgments}
 
     The authors are indebted to K. Kato for giving instruction in the 
method for numerical computations of Feynman parameter integrals \cite{kato}.   
This  work is supported in part by the Grant-in-Aid for Scientific
Research (No. 08640357, No. 08640400) and the Grant-in-Aid for 
Scientific Research on Priority Areas (Physics of $CP$ Violation, 
No. 09246211) from the Ministry of Education, Science and
Culture, Japan.

\newpage 

\begin{table}
\begin{center}
\begin{tabular}{c c c c c}
\hline
    & (i.a) & (i.b) & (ii.a) &(ii.b) \\
\hline
 $\m2$ (GeV) & 100 & 100 & 200 & 200 \\
 $\tb$  & 2  & 10  & 2 & 10  \\
\hline
\end{tabular}
\end{center}
\caption{The values of $\m2$ and $\tb$ for curves
    (i.a)--(ii.b) in Figs. 2 and 3.  
}
\label{tab1}
\end{table}
\begin{table}
\begin{center}
\begin{tabular}{c c c c c}
\hline
 $\mwi$ (GeV) & 133 & 274 &  &  \\
 $\mxj$ (GeV)  & 85  & 146  & 203 & 278  \\
\hline
\end{tabular}
\end{center}
\caption{The mass spectra of the charginos and 
neutralinos in Figs. 4 and 5.  
}
\label{tab2}
\end{table}

\pagebreak 

\begin{figure}
\begin{center}
     \setlength{\unitlength}{1mm}
     \begin{picture}(70,90)

\multiput(3,5)(0,40){2}{
  \multiput(5,15)(5,0){3}{\bezier{200}(0,0)(1.25,1.5)(2.5,0)}
  \multiput(7.5,15)(5,0){3}{\bezier{200}(0,0)(1.25,-1.5)(2.5,0)}
  \multiput(40,25)(5,0){3}{\bezier{200}(0,0)(1.25,1.5)(2.5,0)}
  \multiput(42.5,25)(5,0){3}{\bezier{200}(0,0)(1.25,-1.5)(2.5,0)}
  \multiput(40,5)(5,0){3}{\bezier{200}(0,0)(1.25,1.5)(2.5,0)}
  \multiput(42.5,5)(5,0){3}{\bezier{200}(0,0)(1.25,-1.5)(2.5,0)}
  \put(20,15){\line(2,1){20}}
  \put(20,15){\line(2,-1){20}}
  \put(40,25){\line(0,-1){20}}
  \put(58,23.5){$W$}
  \put(58,3.5){$W$}
}
  \put(0,58.5){$Z, \gamma$}
  \put(4,18.5){$Z$}

\multiput(3,5)(0,40){1}{
     \put(28.5,64){$\wi$}
     \put(28.5,45){$\wj$}
     \put(42,55){$\xk$}
     \put(28.5,24){$\xi$}
     \put(28.5,5){$\xj$}
     \put(42,15){$\wk$}
}
     \put(21.25,45){(a)}
     \put(21.5,5){(b)}

    \end{picture}
\end{center}
\caption{One-loop diagrams mediated by charginos and neutralinos 
which induce $CP$-odd couplings 
              for $WWZ$ and $WW\gamma$.}
\label{fig1}
\end{figure}
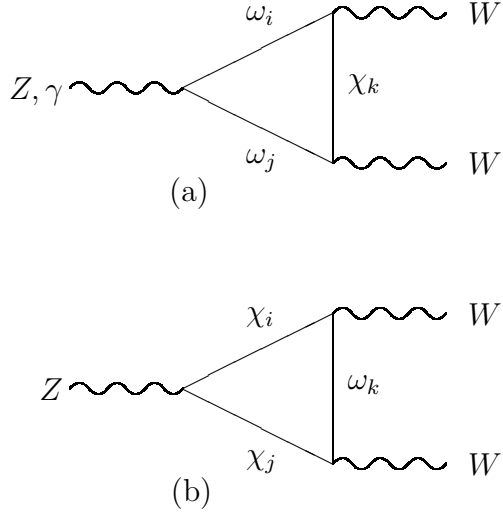

\pagebreak

\begin{figure}
\vspace{8cm}
\includegraphics{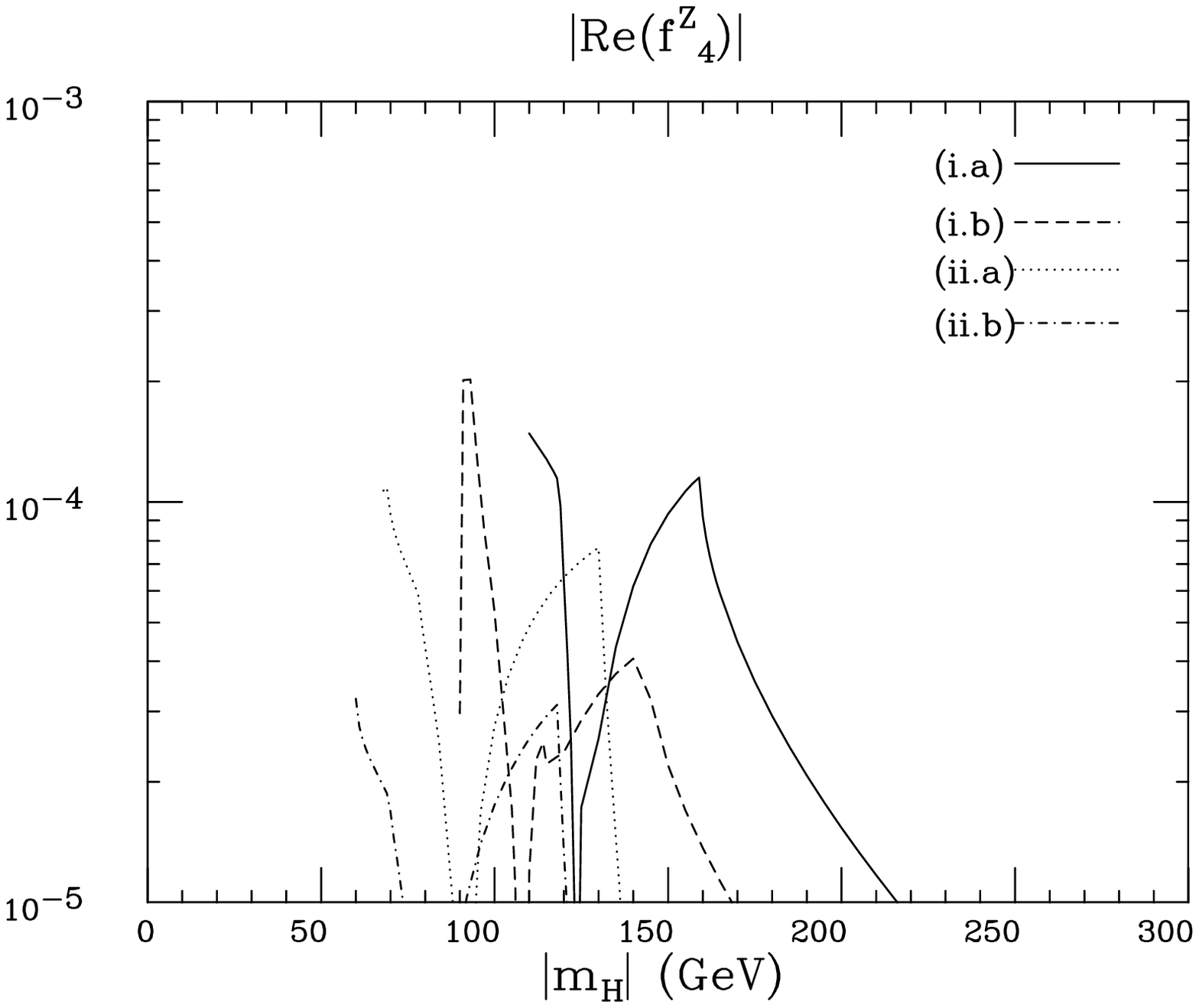}
\begin{center}
Fig. 2(a)
\end{center}
\vspace{9cm}
\includegraphics{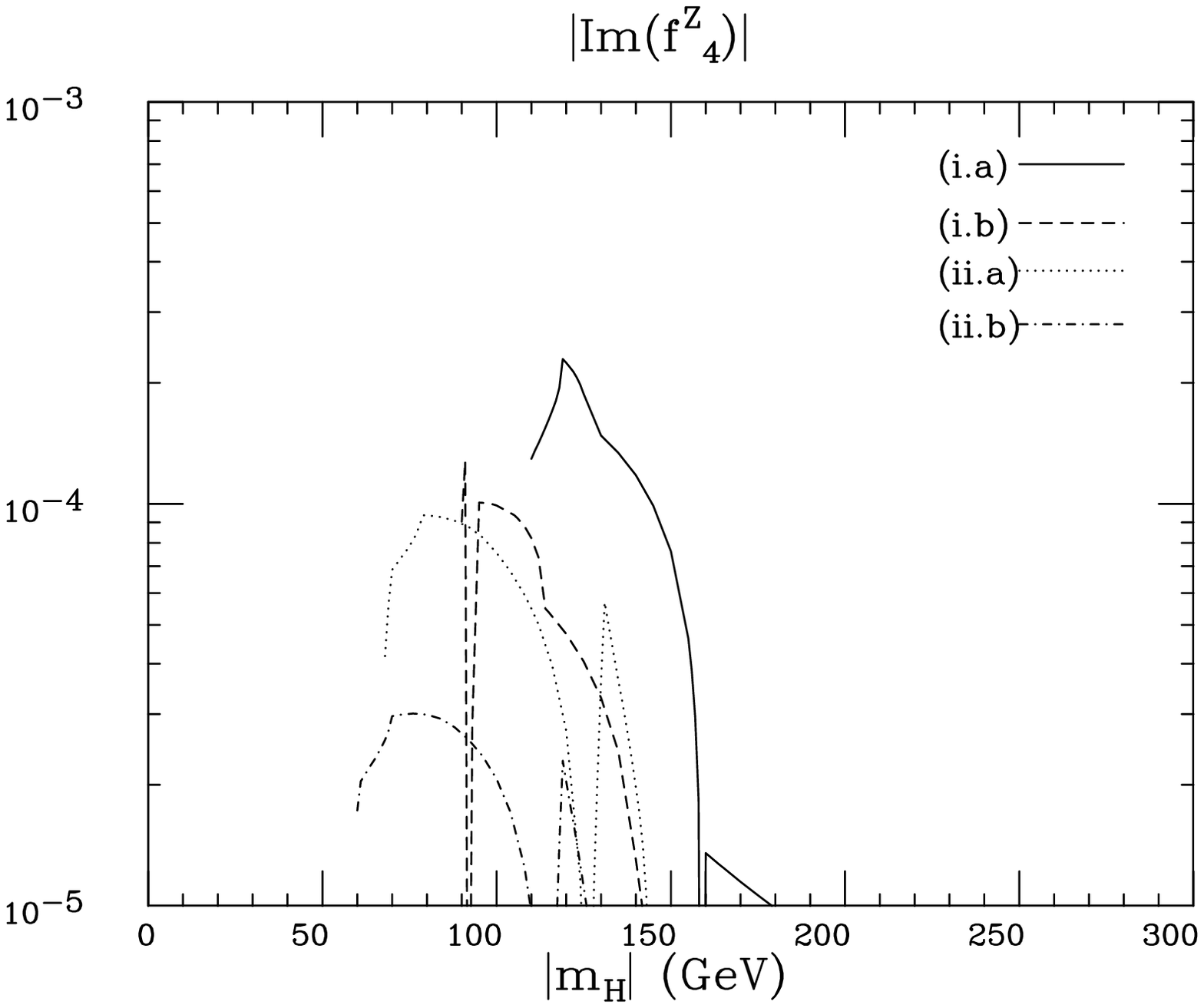}
\begin{center}
Fig. 2(b)
\end{center}
\end{figure}
\begin{figure}
\vspace{8cm}
\includegraphics{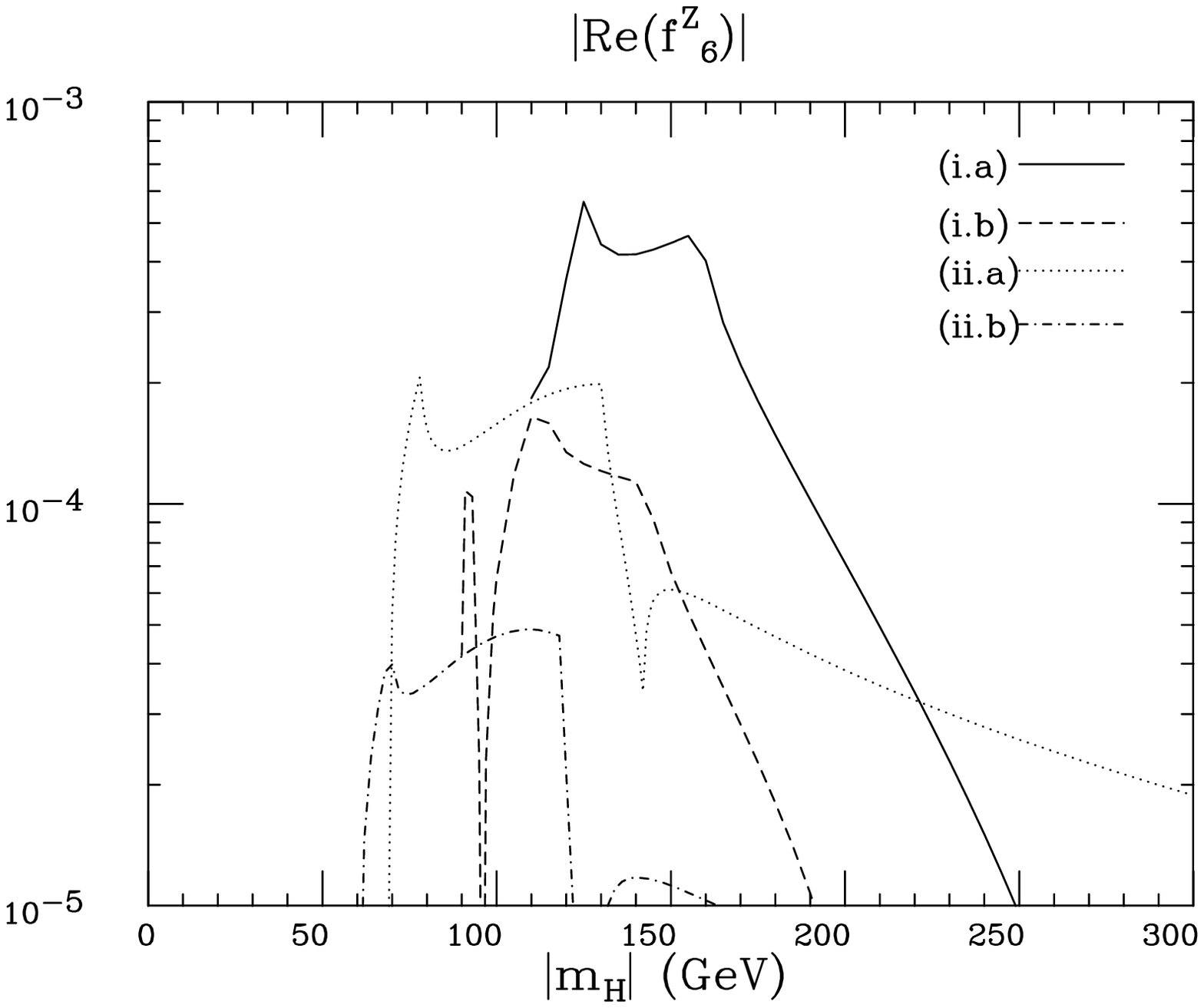}
\begin{center}
Fig. 2(c)
\end{center}
\vspace{9cm}
\includegraphics{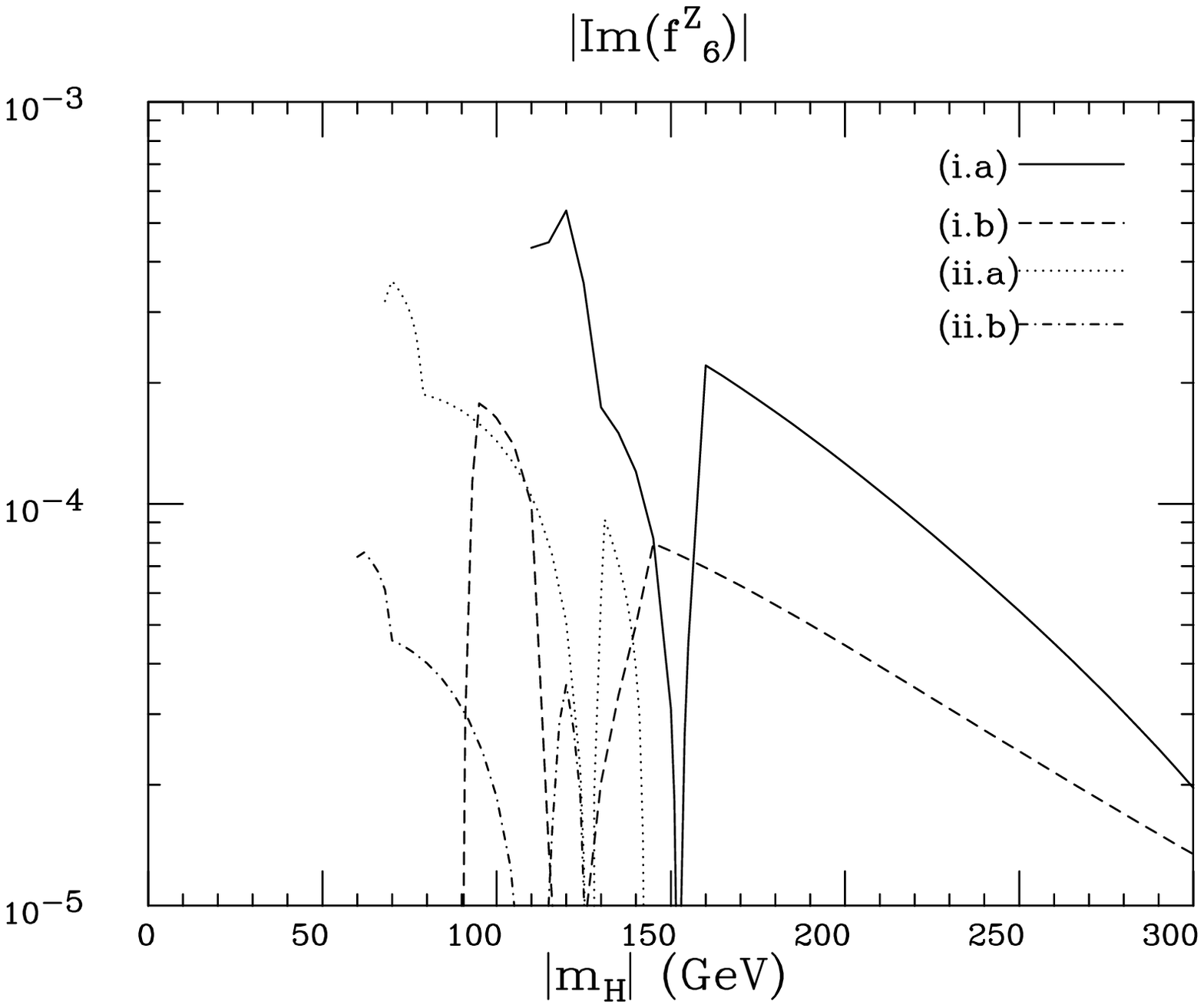}
\begin{center}
Fig. 2(d)
\end{center}
\caption{The absolute values of the real and imaginary 
parts of $f^Z_4$ and $f^Z_6$  as functions of $|\mH|$ 
for $\theta=\pi/4$ at $\protect \sqrt{q^2}=200$ GeV.   
Four curves (i.a)--(ii.b) correspond to the four sets of parameter values 
given in Table 1.  
(a) Re($f_4^Z$), (b) Im($f_4^Z$),  (c) Re($f_6^Z$), (d) Im($f_6^Z$).} 
\label{fig2}
\end{figure}

\pagebreak

\begin{figure}
\vspace{8cm}
\includegraphics{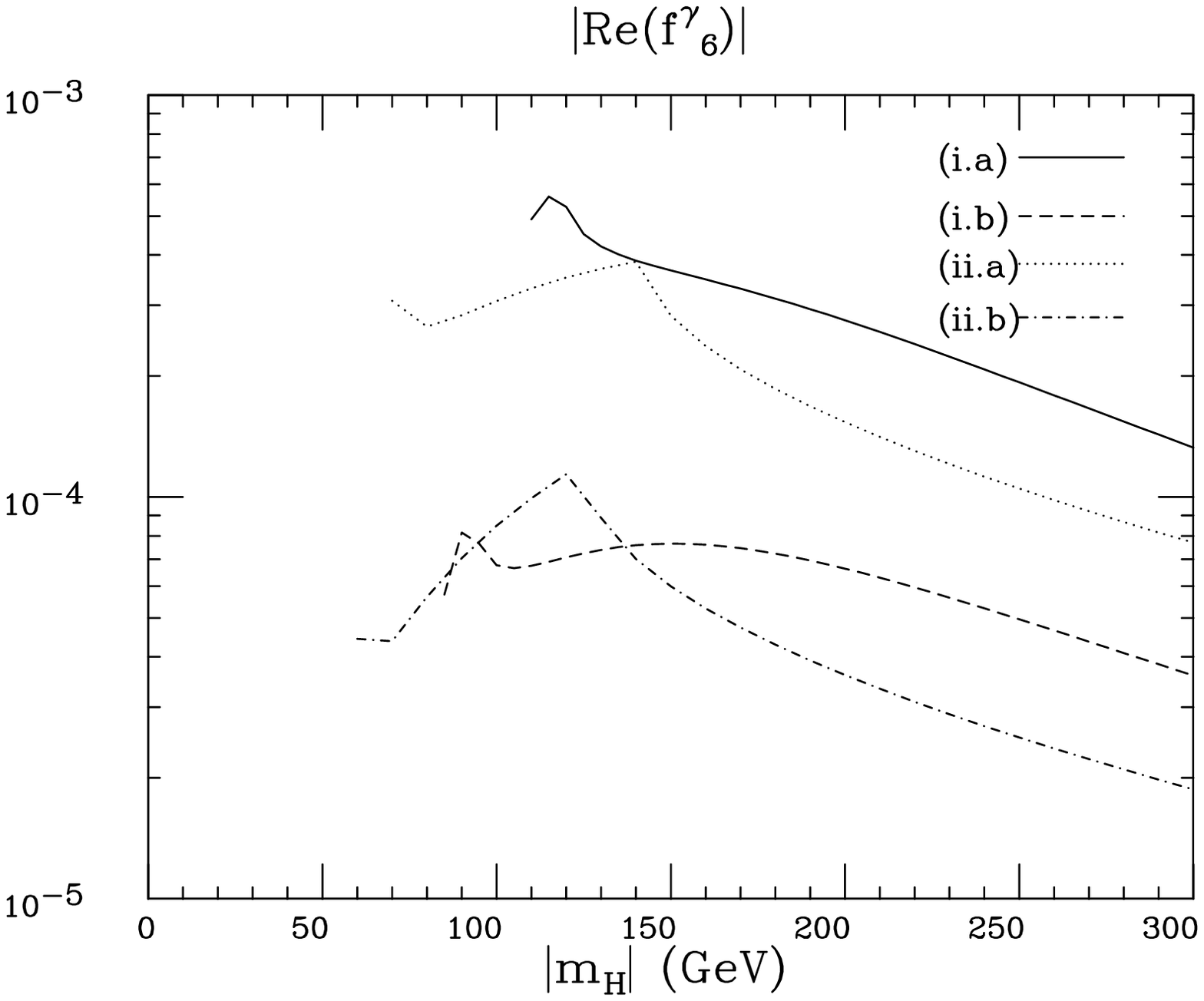}
\begin{center}
Fig. 3(a)
\end{center}
\vspace{9cm}
\includegraphics{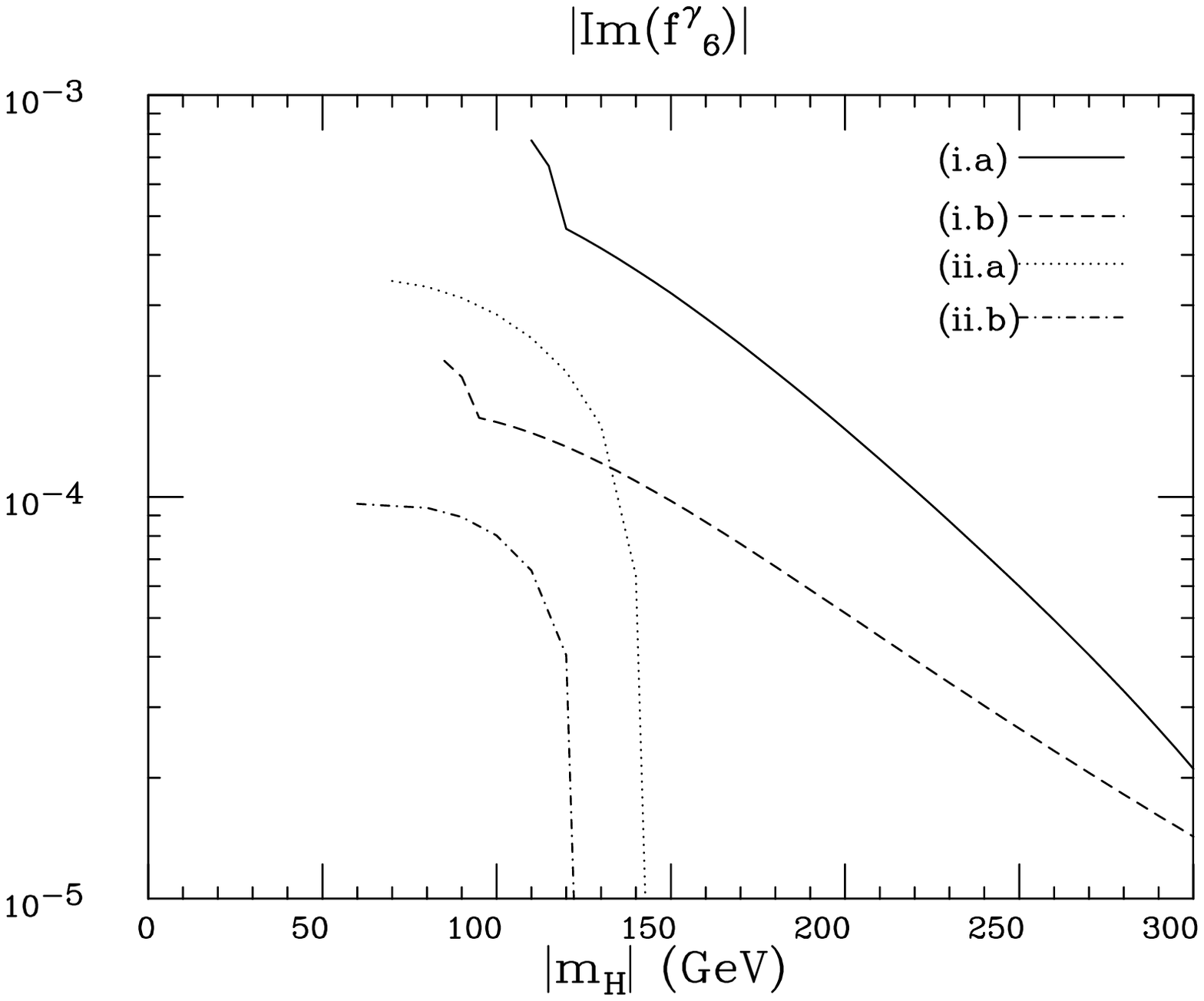}
\begin{center}
Fig. 3(b)
\end{center}
\caption{The absolute values of the real and imaginary 
parts of $f^\gamma_6$  as functions of $|\mH|$ 
for $\theta=\pi/4$ at $\protect \sqrt{q^2}=200$ GeV.  
Four curves (i.a)--(ii.b) correspond to the four sets of parameter values 
given in Table 1.
(a) Re($f_6^\gamma$), (b) Im($f_6^\gamma$). }
\label{fig3}
\end{figure}

\pagebreak 

\begin{figure}
\vspace{8cm}
\includegraphics{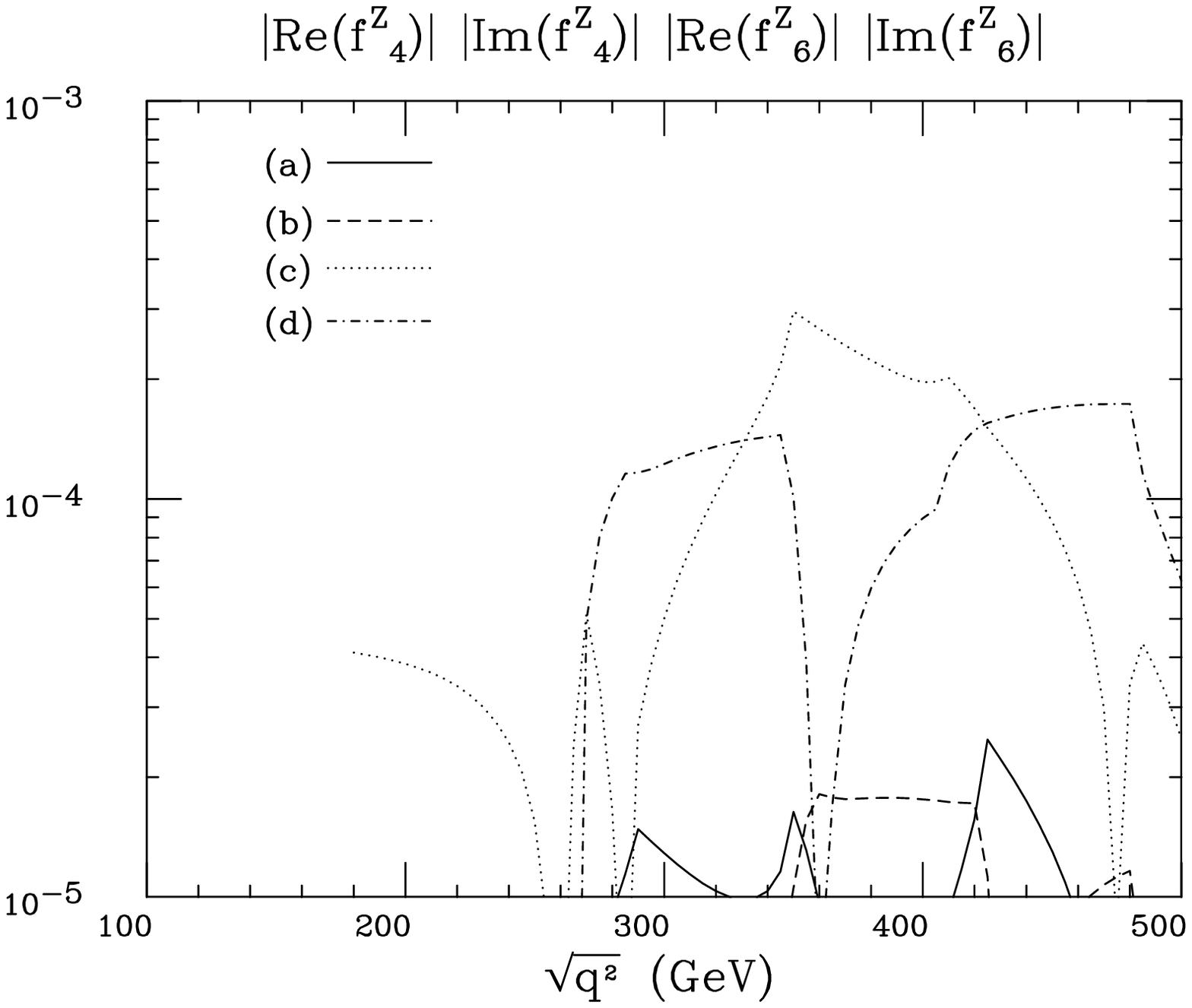}
\begin{center}
Fig. 4
\end{center}
\caption{The absolute values of the real and imaginary 
parts of $f^Z_4$ and $f^Z_6$  as functions of $\protect \sqrt{q^2}$ 
for $\tb=2$, $\m2=200$ GeV, $|\mH|=200$ GeV, and $\theta=\pi/4$.   
(a) Re($f_4^Z$), (b) Im($f_4^Z$), (c) Re($f_6^Z$), (d) Im($f_6^Z$).} 
\label{fig4}
\end{figure}
\begin{figure}
\vspace{8cm}
\includegraphics{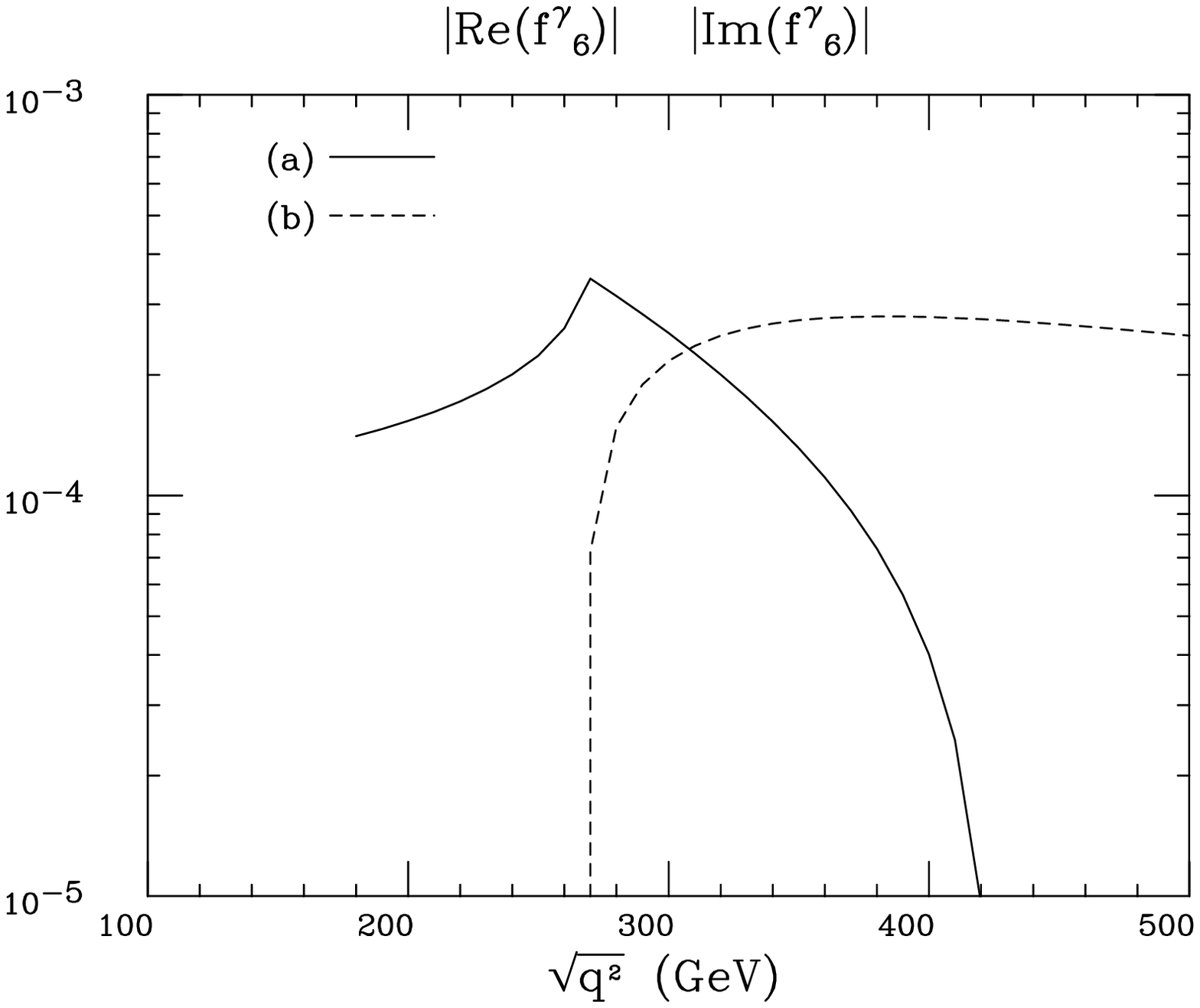}
\begin{center}
Fig. 5
\end{center}
\caption{The absolute values of the real and imaginary 
parts of $f^\gamma_6$  as functions of $\protect \sqrt{q^2}$ 
for $\tb=2$, $\m2=200$ GeV, $|\mH|=200$ GeV, and $\theta=\pi/4$.   
(a) Re($f_6^\gamma$), (b) Im($f_6^\gamma$).}
\label{fig5}
\end{figure}
\end{document}